# High Sensitivity Transient Holographic Full-Polarization Microscopy

**Julia A. Gessner[1*], Philipp Kollenz[1], Godwin Maliyekkal[1], Milena R. Neubert[1], Garrett May[1], Lukas Becker[1,2], Nasrin Solhtalab[1], Tiago Buckup[1], Felix Deschler[1]**

[1]*Institute for Physical Chemistry, Ruprecht-Karls-University Heidelberg, Heidelberg, Germany*
[2]*Karlsruhe Institute of Technology, Karlsruhe, Germany*

**julia.gessner@pci.uni-heidelberg.de*

**Abstract:**

The polarization state of light provides direct access to ultrafast electronic, magnetic, and structural dynamics in materials. Transient wide-field microscopy with simultaneous amplitude, phase, and polarization sensitivity enables spatially resolved measurements of these dynamics, but its sensitivity is often limited by technical noise and background fluctuations. Here, we demonstrate transient holographic full-polarization microscopy with substantially reduced instrumental noise and sensitivity approaching the shot-noise limit. We compare holographic geometries using one, two, and four reference beams and find that the single-reference configuration provides the most favourable shot-noise-limited sensitivity for the derived polarization observables. Combined with two-camera detection, it simultaneously captures both orthogonal polarization components without distributing reference photons among multiple carriers. A self-referenced noise-reduction framework combining singular value decomposition and background subtraction further suppresses correlated noise. For spatially averaged regions, the baseline noise is reduced by factors of 3–60, enabling detection of weaker transient responses at lower excitation densities and in weakly absorbing materials.

## 1. Introduction

The advancement of optoelectronic and spintronic technologies to nanoscale and heterogeneous structures requires spatially resolved measurements of ultrafast electronic and magnetic dynamics in order to understand and optimize the performance of emerging functional materials[1–4]. In pump-probe microscopy, these dynamics are spatiotemporally mapped through transient changes induced by a weak probe pulse. To minimize sample damage, avoid Auger and state-filling effects, and maintain excitation densities relevant to device operation, the pump fluence is typically kept low, resulting in transient intensity changes in the probe, which are often several orders of magnitude smaller than the total probe intensity. Therefore, achieving high detection sensitivity is a central challenge in ultrafast microscopy[5–8]. While the fundamental sensitivity is determined by the shot noise of the detected photons, practical measurements are often dominated by correlated fluctuations arising from laser intensity noise, optical drifts, and sample motion [9]. Balanced detection efficiently suppresses common-mode intensity fluctuations in single-point measurements[10,11], but is less effective at suppressing spatially varying correlated noise in wide-field microscopy. Such spatially structured fluctuations can remain despite balanced detection and can therefore limit the sensitivity of transient wide-field imaging.

For phase-resolved measurements, these limitations are particularly challenging because a weak transient response must be extracted while recording both the amplitude and phase of the optical field. Interferometric detection provides a direct route to this information by encoding the complex probe optical field in the interference between the probe and reference beam. Off-axis holography enables the simultaneous retrieval of amplitude and phase through off-axis interference and has been widely used for wide-field and polarization-sensitive imaging[12– 14]. Its application to transient ultrafast microscopy has further enabled spatially resolved measurements of the transient optical responses[15–17]. In our previous work[18], this approach was extended to a wide-field ultrafast chiroptical microscope capable of simultaneously measuring the transient absorption, circular and linear birefringence, and linear and circular dichroism. However, the multiplexed holographic geometry requires multiple reference beams, which introduce additional constraints on the detected photon budget and detection sensitivity.

Here, we simplified the holographic geometry to a single-reference configuration, thereby reducing the multiplexing-related sensitivity penalty while retaining access to the full complex optical response. For transient measurements, we implemented a single-reference geometry using two cameras, allowing two orthogonal polarization components to be detected simultaneously. We systematically compared three holographic configurations and evaluated their fundamental and experimentally observed noise performances.

To suppress residual correlated noise, we introduce a self-referenced numerical correction based on singular value decomposition (SVD)[19]. The noise basis was derived directly from signal-free regions and negative-delay frames within the acquired dataset, allowing correlated fluctuations to be identified from the statistical properties of the measurement without requiring a separate noise-only acquisition step. We further characterized the residual noise using standard deviation analysis, power spectral density (PSD), principal component analysis (PCA), and the effective number of statistically independent pixels.

We found that the combination of background subtraction and self-referenced SVD correction provides the strongest suppression of correlated noise, bringing the measured fluctuations closer to the Poissonian shot-noise reference. Because SVD correction is self-referenced and relies only on the acquired data, the approach is broadly applicable to wide-field transient microscopy techniques beyond holographic imaging.
The resulting improvement in sensitivity provides a pathway toward wide-field transient measurements of weak optical responses in low-dimensional and nanoscale materials, organic semiconductor thin films, and at low excitation densities that are relevant to device operation.

## 2. Experimental methods

### *2.1 Setup*

The time-resolved characterization of the electronic and magnetoptical responses requires full reconstruction of the transient polarization ellipse of the probe beam (Fig. 1d). To identify the configuration with the highest sensitivity, we compared three experimental layouts and implemented multiple noise-subtraction methods. As shown in Fig. 1a and c, the three setups differ mainly in their detection architectures.

We sent 35 fs laser pulses at a repetition rate of 1 kHz and wavelength of 800 nm from an amplified Ti:sapphire laser (Astrella, Coherent) to two home-built Noncollinear Optical Parametric Amplifiers (NOPAs) for tuning the wavelength of the pump and probe pulses. Both the pump and probe pulses had a set bandwidth of 10 nm. As materials exhibiting strong spin-orbit coupling (SOC) can be magnetized via interaction with circularly polarized light[20–22], we introduce a variable phase retarder (LCC1511-A, Thorlabs) in the pump path to precisely adjust its polarization and efficiently induce spin polarization in the sample. The pump was then focused on the sample from the back of the objective lens (Olympus, UPLAPO, NA = 1.2,

x60 magnification) used in the widefield microscope. In this way, the pump illumination can be set in a diffraction-limited or wide-field (by adding a lens and iris before the objective, see Fig.1a) configuration.

The probe beam was loosely focused on the sample using a lens, and the transmitted light was collected by the objective. Using a 90:10 (T/R) beam splitter before the sample, we used a portion of the probe light to create the reference beam, which interferes with the probe at a small angle on the camera, enabling off-axis holographic detection. In this manner, we can retrieve the amplitude and phase of the probe beam[14,18,23]. We set the polarization of both the probe and reference beams to 45° using a half-wave plate.

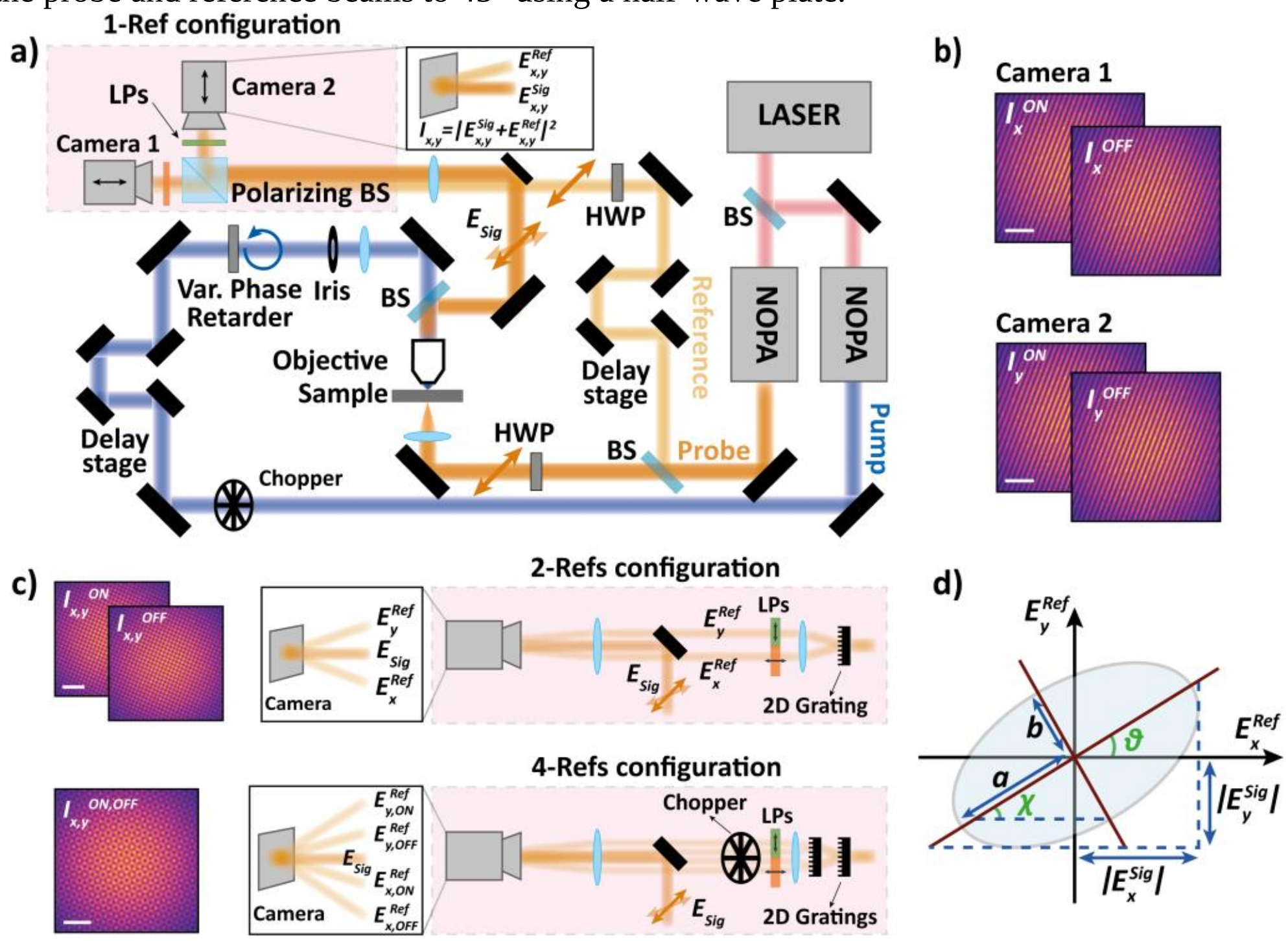

Figure 1: a) Schematic of transient holographic full-polarization microscopy in the single-Reference (1-Ref) geometry. A polarizing beam splitter (BS) separates the s- and p-polarized probe and reference beams and directs them onto two separate cameras. LP: linear polarizer; HWP: half-wave plate. b) Acquired images from the two cameras in the 1-Ref configuration. The ON, OFF, $x$, and $y$ polarization components are acquired separately. c) Double-Reference (2-Refs, top) and Quadruple-Reference (4-Refs, bottom) geometries. In the 2-Refs configuration, the $x$ and $y$ components of the electric field are acquired through polarization multiplexing, while ON and OFF images are recorded separately. In the 4-Refs configuration, all four polarization components ($x$ and $y$, each for ON and OFF) are obtained from a single camera acquisition. d) Polarization ellipse of the probe beam illustrating the polarization parameters $\vartheta$ and $\chi$. Scale bars in all images correspond to 5 μm.

In the first studied detection configuration (Fig. 1a), the probe and reference beams were split into their orthogonal polarization components (x and y, or // and ⊥) with a polarizing beam splitter and imaged onto two separate cameras (Cameras 1 and 2). Additional polarization filtering is provided by placing a linear polarizer in front of each camera. The orthogonality and balanced intensity of the two reference beams were confirmed by rotating the linear polarization of the probe and observing selective interference with the *s*- and *p*-polarized references and equal interference at 45°, which is consistent with mutually orthogonal reference polarizations. Although differences in the gain or spatial response between the two detection channels can introduce a static imbalance between the measured polarization components, this effect largely cancels in the transient measurements: pump-ON and pump-OFF frames were recorded for each polarization channel, and relative changes were computed, while additive background contributions were removed by subsequent noise subtraction.

For the transient measurements shown in Fig. 5, we used two acA720-520um Basler cameras with a 540 × 540-pixel acquisition region and 6.9 µm pixel size. The corresponding field of view (FOV) is 33 × 33 µm$^2$. For the noise-characterization measurements shown in Figs. 2–4, which will be discussed in 4.2 and 4.3, we used a single ace 2 a2A2448-75umPRO Basler camera with 1021 × 1021-pixel acquisition regions and 2.7 µm pixel size, corresponding to less than half of the full sensor dimensions. Here, the FOV was 25 × 25 µm$^2$ for all geometries tested.

Complex fields were independently reconstructed from the two polarization channels. In the absence of pump-induced changes, the reconstructed fields in the two polarization channels remained unchanged between the pump-ON and pump-OFF conditions, yielding a net transient signal of zero. However, if the pump modifies the material properties, the amplitude and/or phase relation of $E_x^{sig}$ with $E_x^{ref}$ and $E_y^{sig}$ with $E_y^{ref}$ becomes unequal, resulting in a nonzero net signal.

As shown in Fig.1d, the detection of the amplitude $E_{x,y}$ and phase $\varphi_{x,y}$ of the measured signal allows retrieval of the polarization-ellipse orientation $\vartheta$ related to the circular birefringence

$$\vartheta = \frac{1}{2}\arctan\left(\frac{2|E_x||E_y|\cos(\varphi_y - \varphi_x)}{|E_x|^2 - |E_y|^2}\right) \tag{1}$$

as well as the ellipticity angle $\chi$, which quantifies circular dichroism (CD):

$$\chi = \frac{1}{2}\arcsin\left(\frac{2|E_x||E_y|\sin(\varphi_y - \varphi_x)}{|E_x|^2 + |E_y|^2}\right) \tag{2}$$

Where $\varphi = \varphi_y - \varphi_x$ denotes the relative phase between the two orthogonal polarization components. Moreover, we can retrieve the linear dichroism ($LD$), given by the normalized difference in the transmitted intensity for two orthogonal polarization directions, and the static absorbance $A$ (or transmission $T$), which is given by the sum of the two intensities normalized over the incident intensity $E_{in}^2$:

$$1 - A = T = \frac{E_x^2 + E_y^2}{E_{in}^2} \tag{3}$$

In order to obtain the transient signals, we subtract the respective signals in presence and in absence of the pump to obtain $\Delta\vartheta$, $\Delta\chi$, $\Delta\varphi$ and $\Delta T$.

The single-reference configuration (1-Ref) requires the acquisition of four separate images distributed across the two cameras, as each camera records two sequential images in the presence (pump ON) and absence of the pump (pump OFF) (Fig. 1b). The image acquisition was synchronized with the pump chopper, allowing unambiguous identification of the pump-ON and pump-OFF images. The pump beam was modulated at 100 Hz, resulting in five consecutive pump-ON pulses, followed by five pump-OFF pulses per modulation cycle. At each pump–probe delay position, 500 frames were acquired with the pump ON, and an additional 500 frames with the pump OFF. The modulation frequency and number of acquired frames were chosen to make efficient use of the available camera frame rate, while maximizing the usable dynamic range of the cameras.

Figure 1c shows the two additional geometries investigated in this study: the double reference (2-Refs), as used in Ref. [18] and quadruple reference (4-Refs) configuration. As shown in the upper panel, a 2D transmission grating (MS-561, Holo-OR) is introduced to generate two reference beams from two of the four first-order diffracted beams. Because only the selected diffraction orders are used as reference beams, the remaining diffracted power does not contribute to the detected interference signals, resulting in additional optical losses in the multiplexed configurations. The grating also controls the relative pulse fronts of the probe and reference beams, which is important for maintaining efficient interference for femtosecond pulses[24]. By contrast, the 1-Ref geometry does not employ a diffraction grating and operates at a small angle between the probe and reference beams. The resulting temporal walk-off is therefore reduced compared with larger-angle geometries, while the use of a 10-nm bandpass filter before the polarizing beam splitter further limits the spectral bandwidth reaching the cameras. Under these conditions, the temporal coherence requirements for efficient interference were relaxed, allowing stable holographic detection without active pulse-front matching. Two cross-polarized linear polarizers were used to set the polarization of the references to *p*-polarization ($E_x^{ref}$) and *s*-polarization ($E_y^{ref}$), respectively, enabling the detection of the probe polarization ellipse. To acquire transient signals, two sequential images were recorded for the pump-ON and pump-OFF states.

In the fully multiplexed configuration shown in the lower panel of Fig. 1c, the complete transient polarization state of the probe was encoded by a single camera exposure. To achieve this, two 2D transmission gratings are introduced to generate two pairs of reference beams. Using two cross-polarized linear polarizers, the two pairs were set to *p* ($E_x^{ref}$) and *s* ($E_y^{ref}$) polarizations. A chopper was then used to synchronize one reference from each pair alternately with the pump, whose repetition frequency was half that of the probe. This generates four reference states $E_{x,ON}^{ref}$, $E_{x,OFF}^{ref}$, $E_{y,ON}^{ref}$ and $E_{y,OFF}^{ref}$, which are simultaneously recorded in a single camera exposure.

Increasing the number of reference beams introduces additional complexity into the detection geometry as it becomes increasingly challenging to ensure homogeneous and well-defined interference between the probe and all reference beams across the camera sensor. In multiplexed configurations, multiple interference terms share the same camera exposure and dynamic range of the detector. Consequently, the available dynamic range per interference term is reduced compared with the sequential acquisition used in the single-reference scheme. Configurations with fewer simultaneously recorded channels provided a higher effective dynamic range for each interference term. This aspect is discussed in Section 3.4.

## 3. Theoretical background

### *3.1 Noise sources*

In the following section, we analyze the dominant noise sources that limit the signal-to-noise ratio in wide-field transient holographic microscopy. The measured signal can be expressed as [9,25]

$$S_{meas} = S_{real} + \sigma \tag{4}$$

where $S_{real}$ is the true physical signal arising from the material response to photoexcitation and $\sigma$ is the total measurement noise. For the purpose of estimating the relative contributions of these noise sources, we use the following approximate root-sum-square noise budget:

$$\sigma = \sqrt{{\sigma_{pump}}^2 + {\sigma_{probe}}^2 + {\sigma_{readout}}^2 + {\sigma_{dark}}^2 + {\sigma_{shot}}^2 + {\sigma_{drift}}^2 + {\sigma_{holo}}^2} \tag{5}$$

Here, $\sigma_{shot}$ denotes the photon shot noise, while $\sigma_{pump}$ and $\sigma_{probe}$ represent the noise originating from fluctuations in the pump and probe beams, including pulse-to-pulse energy variations and pointing jitter. The camera readout and dark noise are denoted as $\sigma_{readout}$ and $\sigma_{dark}$, respectively. The term $\sigma_{drift}$ accounts for spatially correlated noise arising from sample motion or beam drifts, which is particularly detrimental when measuring charge-carrier diffusion on sub-picosecond timescales in samples with complex spatial morphology. Finally, $\sigma_{holo}$ describes the noise associated with imperfect interference between the probe and reference beams, for example, owing to slight optical misalignments. This decomposition is an approximate noise budget rather than a decomposition into statistically independent noise sources. In practice, several contributions can be correlated. These covariance terms are not explicitly included in Eq. (5); instead, the equation provides a convenient estimate of the magnitude of the individual contributions to the overall experimental noise.

In our transient measurement configuration, the camera readout and dark noise are negligible (<0.02 counts RMS for 500 exposures) compared with the photon shot noise. Therefore, the fundamental sensitivity limit is set by shot noise, which follows Poisson statistics with a standard deviation of:

$$\sigma_{shot} = \sqrt{N} \tag{6}$$

where $N$ is the number of detected photoelectrons. Consequently, the relative shot noise scales as $1/\sqrt{N}$, making it the fundamental noise floor. Therefore, the objective of the detection scheme is to suppress the dominant instrumental noise contributions, particularly $\sigma_{pump}$, $\sigma_{probe}$, $\sigma_{drift}$ and $\sigma_{holo}$, such that the total experimental noise approaches the fundamental shot noise limit.

### *3.2 Zernike polynomials and background subtraction*

Zernike polynomials provide an orthogonal basis for describing smooth spatial variations over a circular aperture, and are widely used for wavefront and aberration analyses[26,27]. This makes them suitable for modelling slowly varying background contributions in spatially resolved measurements, such as defocus, astigmatism, or probe-reference wavefront mismatch. The background is represented as a weighted sum of the Zernike modes:

$$B(x,y) \approx \sum_{k=1}^{M} a_k Z_k\,(x,y) \tag{7}$$

Where $Z_k(x,y)$ are the Zernike polynomials, $a_k$ are coefficients, and $M$ is the number of modes included in the expansion. The coefficients were obtained by least-squares fitting to the regions of the measured image, which were assumed to contain only background contributions.

The low-order modes describe progressively more complex spatial variations from a constant offset (piston) to tilt, defocus, astigmatism, and higher-order structures. However, static spatial distortions do not account for all noise and background contributions present in the differential measurements. After differential acquisition and Zernike correction, spatially correlated fluctuations remain, extending over multiple pixels and varying from frame to frame. Such correlations may arise from various instrumental and experimental fluctuations, including variations in the illumination, detector response, and optical or mechanical stability. Because they affect multiple pixels simultaneously, they are not efficiently suppressed by spatial averaging, and can contribute substantially to residual noise. Moreover, because the Zernike correction is based only on spatial shape, it does not intrinsically distinguish between an unwanted background and a genuine low-spatial-frequency signal. Signal components that overlap with the selected Zernike modes may therefore be partially removed during correction.

These limitations motivate a complementary data-driven approach to suppress the remaining correlated noise. While the Zernike correction assumes a predefined spatial basis, our approach based on SVD derives the dominant spatial modes directly from the experimental data. By constructing this basis from regions or time points where no signal is expected, SVD can characterize the correlated fluctuations without imposing a specific spatial functional form. The resulting noise modes were then fitted to the measured background and removed from the full dataset.

### *3.3 SVD-based correlated noise removal*

To remove correlated noise patterns from the data, correlated noise removal (CNR) based on SVD was applied[19,28,29]. This method exploits the fact that fluctuations in the probe intensity, detector response, or other experimental parameters can affect multiple pixels simultaneously. For example, a fluctuation in the probe intensity can produce a simultaneous intensity change across pixels with different transmissions, resulting in a characteristic spatial noise pattern. These correlated patterns can be identified from regions and time points where no pump-induced signal is expected and are subsequently removed from the measurement.

The method relies on three assumptions: (1) no transient signal is present at negative pump-probe delays, (2) the spatial regions outside the pumped area contain no signal, and (3) the correlated noise patterns observed in these signal-free regions remain representative of those affecting the pumped region throughout the measurement.

The measured data were arranged into a two-dimensional data matrix $D$, with spatial positions along the rows and pump-probe delay points along the columns. The subset $D_{pre}$, which contains measurements acquired before time zero, was used to determine the dominant spatial noise modes through SVD:

$$D_{pre} = U\, S\, V^T \tag{8}$$

where $U$ contains the spatial eigenmodes, $S$ contains singular values representing their weights, and $V^T$ describes the frame-to-frame variation of the corresponding mode amplitudes over the negative-delay measurements. Although no pump-induced signal was present at negative delays, the correlated noise amplitudes fluctuated from frame to frame because of laser fluctuations, detector variations, and experimental drifts. Only the first $N$ singular vectors corresponding to the largest singular values are retained as the correlated noise basis.

During the measurement, the contribution of these noise modes was determined for each image using the unpumped detector region, hereafter referred to as the edge pixel region. These pixels are located outside the spatial region excited by the pump, and are therefore assumed to contain no pump-induced signals. The selected spatial modes were fitted to the measured signal at the edge pixels using linear least-squares.

$$\beta = \left(U_{edge}^T U_{edge}\right)^{-1} U_{edge}^T D_{edge} \tag{9}$$

where $U_{edge}$ is the subset of spatial components corresponding to the edge pixels and $D_{edge}$ is the measured signal at the same pixels. The fitted coefficients $\beta$ describe the contribution of each correlated noise mode at a given time point.

The full detector-wide noise contribution was reconstructed as:

$$D_{noise} = U_N \cdot \beta \tag{10}$$

Where $U_N$ is the selected spatial noise basis, and subtracted from the original data:

$$D_{denoised} = D - D_{noise} \tag{11}$$

Unlike the Zernike approach, which uses a predefined spatial basis, SVD derives the noise modes directly from the measured data, and can therefore capture correlated spatial fluctuations with an arbitrary spatial structure. Importantly, the procedure is self-referenced; both the noise basis and frame-by-frame noise amplitudes are obtained from signal-free regions and time points within the same measurement. Consequently, no separate noise-only acquisition is required because the noise basis is extracted from signal-free regions and negative-delay measurements within the same dataset.

### *3.4 Shot-noise for 1-Ref, 2-Refs and 4-Refs configurations*

Multiplexed off-axis holography enables simultaneous acquisition of multiple off-axis interference carriers within a single exposure[30,31]. Here, a carrier denotes the spatial-frequency component introduced by the angular interference between the probe (object) beam and the reference beam. Therefore, each reference beam generates a distinct spatial carrier in the recorded hologram, allowing the corresponding interference term to be isolated in the Fourier domain. The distribution of available photon counts among different carriers introduces additional considerations for signal-to-noise ratio (SNR) optimization. The shot noise at the camera is governed solely by Poisson statistics, and depends only on the total number of detected photons[32].

Denoting the total photon number by

$$N_{tot} = N_s + \sum_{i=1}^{M} N_{r,i} \tag{12}$$

where $N_s$ is the signal photon number, and $N_{r,i}$ is the photon number in the $i$-th reference beam. Therefore, the camera shot-noise variance is

$$\sigma_{shot}^2 = N_{tot} \tag{13}$$

which is independent of the number of reference beams $M$. Thus, for fixed total counts, the raw camera image exhibits identical shot-noise statistics for single, double, and quadruple references.

However, after Fourier-domain filtering and isolation of a single off-axis sideband, the noise properties in the reconstructed complex image differed between the multiplexing cases. The amplitude of the interference term associated with the $i$-th carrier scale is:

$$A_i \propto 2\sqrt{N_s N_{r,i}} \tag{14}$$

If the total reference photon number $N_r = \sum_i N_{r,i}$ is fixed and evenly distributed across the references $N_{r,i} = N_r/M$, the reconstructed signal amplitude per hologram becomes

$$A_i \propto 2\sqrt{N_s \frac{N_r}{M}} = \frac{1}{\sqrt{M}} 2\sqrt{N_s N_r} \tag{15}$$

The amplitude of each individual interference term decreases with increasing multiplexing order, because the available probe and reference photons are distributed among multiple carriers, whereas the shot-noise contribution is determined by the total detected photon budget. Consequently, for a fixed total photon number, the reconstructed interference amplitude per channel decreases with the number of reference beams, thereby increasing the relative shot-noise contribution to each reconstructed channel. Therefore, the shot-noise-limited SNR decreases with increasing multiplexing order:

$$SNR_M = \frac{1}{\sqrt{M}} SNR_{M=1} \tag{16}$$

Although the raw camera shot noise was identical for the 1-Ref, 2-Refs, and 4-Refs geometries at fixed total photon counts, the reconstructed complex fields did not exhibit the same SNR. Photons allocated to additional carriers contribute to the total detected photon count and hence to the camera shot-noise variance, but not to the selected reconstructed interference term, resulting in an effective increase in the noise relative to the reconstructed signal for each individual channel. This behavior is analogous to balanced-detection schemes in which an additional reference channel is introduced to suppress technical noise but its detected photons nevertheless contribute to the shot-noise limit[33]. The two-camera 1-Ref geometry differs in that both polarization channels contribute to the final observables; therefore, their independent shot-noise contributions must be propagated through the corresponding polarization reconstruction and do not result in a universal $\sqrt{2}$ penalty.

## 4. Results and discussion

### *4.1 Fundamental shot-noise limit of the different geometries*

To establish the fundamental sensitivity limit of the different multiplexing geometries, we first performed numerical simulations (Fig. 2a) based on the experimentally measured data (Fig. S1). Numerical holograms were generated for the different reference configurations, using the same k-space allocation for each geometry to isolate the effect of multiplexing. The 1-Ref and 2-Refs configurations leave additional Fourier-space bandwidth unused and could, in principle, exploit this bandwidth to achieve a higher spatial resolution and reduced spatial correlations. The corresponding shot-noise-limited uncertainty was calculated from the expected photon statistics (see the Supplementary Information for details). As shown in Fig. 2b, the simulations reproduced the theoretical multiplexing-induced shot-noise penalty derived in section 3.4. In the absence of technical noise, the single-reference (1-Ref) configuration provided the lowest shot-noise-limited uncertainty and represented the optimal geometry for maximizing sensitivity.

In practice, however, holographic measurements are often dominated by technical noise sources, including laser fluctuations, background signals, and sample or optical path drifts. Nevertheless, multiplexed configurations can be advantageous in this regime because they enable the simultaneous acquisition of multiple polarization channels and pump states, reducing the number of sequential exposures. Therefore, we experimentally characterized all three geometries in combination with the noise-removal strategies introduced above to assess their performance under realistic measurement conditions and their ability to approach the fundamental shot noise limit.

Noise characterization measurements were performed on a thin methylammonium lead tribromide (MAPbBr3) crystal, which was used as a benchmark sample. The corresponding real- and reciprocal-space images for the 1-Ref, 2-Refs, and 4-Refs configurations are shown in Fig. S1. The measurements were carried out under static conditions, without the pump beam, with both the probe and reference beams generated at a wavelength of 550 nm below the optical resonance of the investigated material, and detected using the same camera. As mentioned above, two cameras were used to simultaneously detect the orthogonal s- and p-polarization components in the 1-Ref configuration. However, for noise characterization, all three configurations were measured using a single camera. This ensures comparable detector characteristics and acquisition conditions across the three configurations. Sequential acquisition does not affect the comparison because these measurements characterize the intrinsic and technical noise of the detection scheme rather than the transient polarization dynamics.

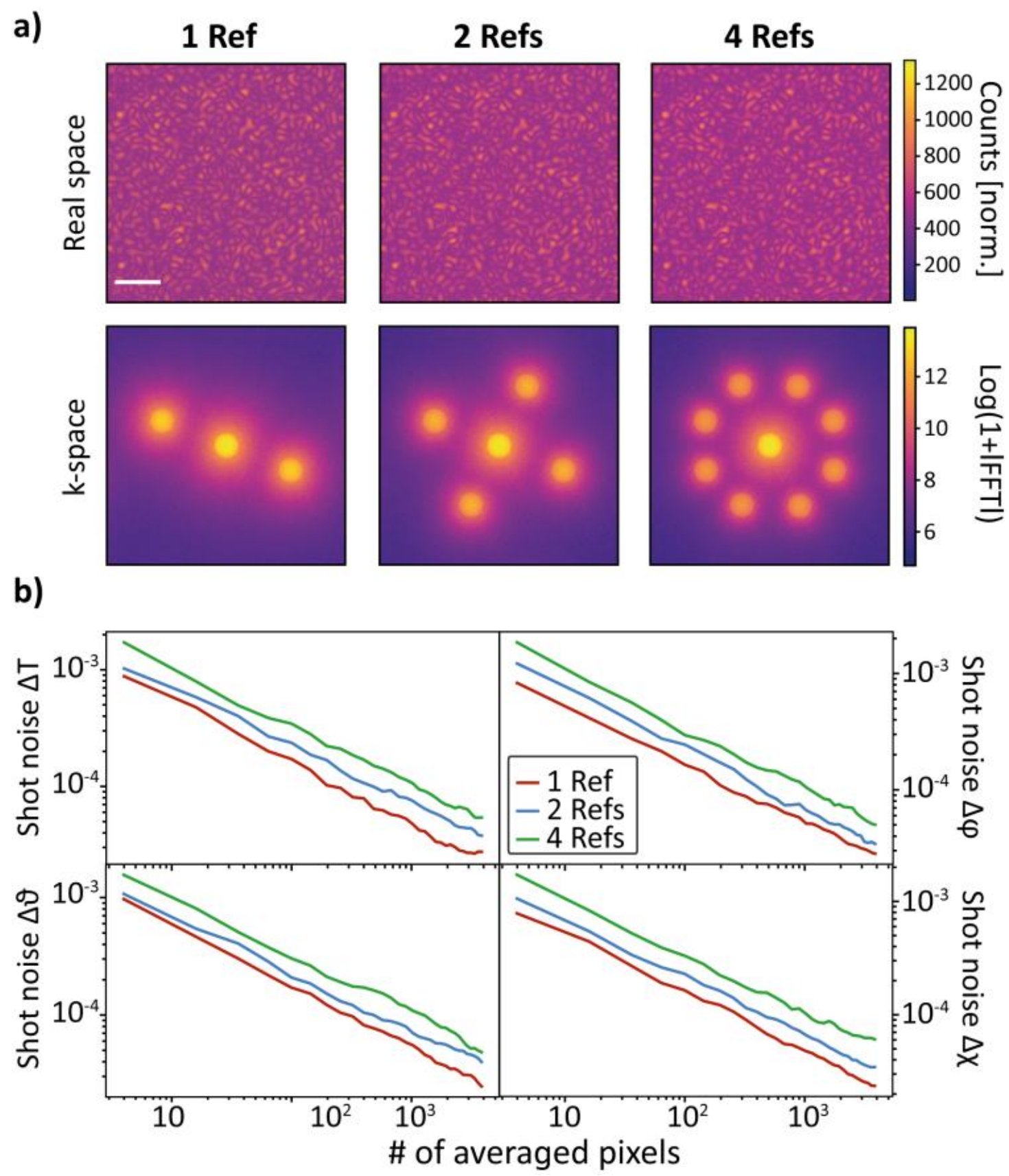

Figure 2: a) Real-space and k-space representations of simulated holograms reproducing the experimentally measured data for the three investigated reference geometries (1-Ref, 2-Refs, 4-Refs). The scale bar corresponds to 5 µm. The k-space representation is shown on a logarithmic scale to account for the stronger central term compared with the other components. b) Shot noise for $\Delta T$, $\Delta\vartheta$, $\Delta\varphi$ and $\Delta\chi$ as a function of the number of spatially averaged pixels, calculated from simulated holograms reproducing the experimental configurations shown in (a).

The measurements were acquired at 120 nominal delay positions using the same total measurement time for all configurations. Although no pump beam was present, the nominal pump-ON/pump-OFF acquisition sequence was retained to reproduce the subtraction and reconstruction procedure used for the transient measurements. At each delay position, 50, 100, and 200 camera frames were acquired per complete acquisition cycle for the 1-Ref, 2-Refs, and 4-Refs configurations, respectively. In the 1-Ref configuration, each cycle comprised four sequential states corresponding to the *s*- and *p*-polarization components under the nominal pump-ON and pump-OFF conditions. In the 2-Refs configuration, the two polarization components are encoded simultaneously in the two reference carriers, whereas the pump-ON and pump-OFF states are acquired sequentially. In the fully multiplexed 4-Refs configuration, both polarization components and pump states are encoded simultaneously. The number of frames was increased with the multiplexing order to compensate for the distribution of the photon budget among the multiple interference carriers.

Prior to averaging, each hologram was phase-corrected to compensate for the slowly varying global phase drifts between acquisitions. Without this correction, phase fluctuations would lead to partial destructive interference during averaging, and consequently degrade, rather than improve, the SNR. Noise subtraction was applied to each frame before averaging.

Small variations in the detected camera counts between configurations were accounted for by normalizing the measurements to their respective detected intensities. Despite this normalization, the removal and reinserting of the two-dimensional diffraction gratings introduced small alignment changes in the optical setup. Therefore, the comparison of the experimentally determined SNRs should be interpreted primarily in terms of the noise-suppression trends obtained for each configuration, rather than as a strict quantitative ranking of their absolute noise performance. The advantage of the 1-Ref geometry in the fundamental shot-noise limit was established by the numerical simulations shown in Fig. 2.

We then applied the noise-removal strategies introduced above to the experimental data and systematically evaluated their performances. The analysis quantifies the residual noise using several complementary metrics, including the standard deviation of the residual fluctuations, power spectral density (PSD) of the noise, eigenvalue spectrum obtained from principal component analysis (PCA), and effective number of independent pixels, which characterizes the degree of spatial noise correlation. Together, these metrics assess both the magnitude and spatial structure of the residual noise, allowing us to determine how effectively different correction strategies suppress technical noise and approach the fundamental shot-noise limit.

In the main text, we focus on the 1-Ref configuration, which is predicted from the simulations to provide the highest sensitivity in the shot-noise limit (Fig. 2). The corresponding results for the 2-Refs and 4-Refs configurations are shown in Figs. S6–S7 and S8–S9, respectively, and provide a comparison of the noise-removal performance across the different multiplexing geometries.

### *4.2 Power spectral density and PCA eigenvalues spectrum*

Next, we compared three noise correction approaches in the 1-ref geometry: CNR, third-order Zernike polynomial correction, and CNR combined with prior background subtraction (CNR+BG), where the background is represented by the zeroth-order Zernike mode. Figure 3a shows a representative image sequence of the detector. The red region (35 pixels wide) denotes the background area used for background characterization and noise correction, whereas the remaining central region defines the signal area used for the subsequent analysis. The following discussion focuses on $\Delta\varphi$ and $\Delta\vartheta$, while the corresponding results for the other observables are provided in Figs.S2-S5.

To evaluate the temporal frequency dependence of noise suppression, we calculated the ratio of the temporal power spectral densities $PSD\ ratio = PSD_{corr}/PSD_{uncorr}$ [34,35]. Values below unity indicate a reduction in the noise power after correction. The temporal frequency is defined with respect to the sampling rate of the reconstructed measurement, rather than the raw camera exposure rate. Thus, the PSD describes how the reconstructed quantity varies over time. Further details of the PSD evaluation are provided in Supplementary Information.

Figures 3b,c show that all correction procedures reduce the temporal noise power over the investigated frequency range. Across nearly the entire frequency range, CNR and CNR combined with background subtraction (CNR+BG) produced lower residual PSD values than the Zernike correction. The PSD ratios remained relatively uniform across the investigated frequency range, indicating that the relative noise-reduction performance of the correction methods does not exhibit a strong temporal-frequency dependence. Nevertheless, the lowest-

frequency components should be interpreted with caution, as their estimation is sensitive to the finite duration of the recorded time series, residual slow drift, nonstationarity, detrending, and spectral leakage. Method-dependent separation remained visible across the investigated frequency range, rather than converging to a common noise floor at higher frequencies. Thus, within the investigated bandwidth, CNR-based methods provide more effective temporal noise suppression than Zernike correction.

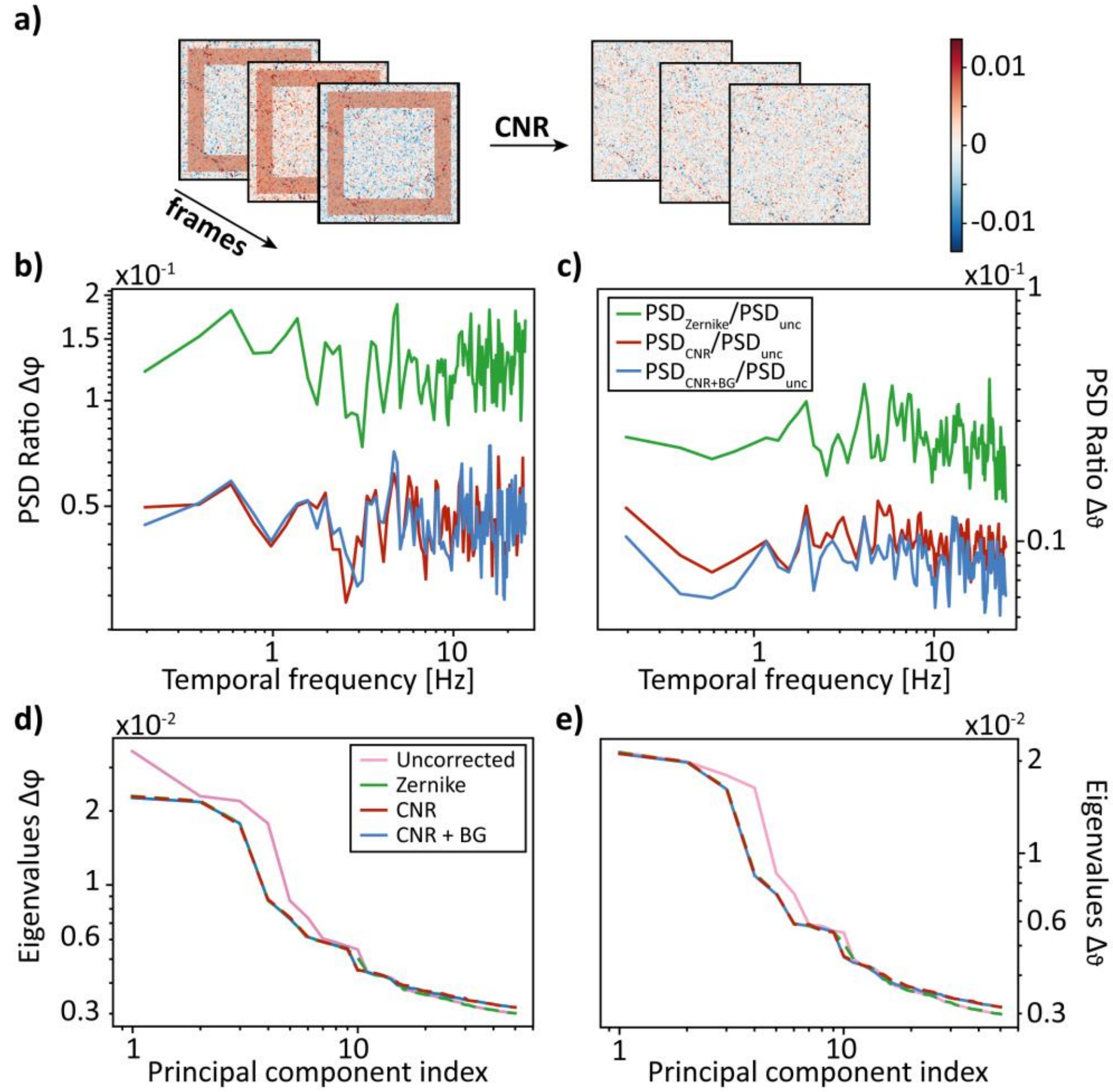

Figure 3: (a) Sequential reconstructed images of the physical observable before (left) and after applying correlated noise removal (CNR) (right). The red frame indicates the background region used as a noise reference for the correction. (b,c) Ratio of the power spectral densities (PSD) as a function of temporal frequency for $\Delta\varphi$ and $\Delta\vartheta$, respectively. Values smaller than one indicate a reduction of the PSD after correction. (d,e) Eigenvalues obtained from PCA as a function of the PCA index for $\Delta\varphi$ and $\Delta\vartheta$, respectively. The similar overall decay of the eigenvalue spectra for the corrected and uncorrected datasets indicates that the correction procedure reduces correlated noise while largely preserving the dominant variance structure of the data.

Whereas the PSD analysis describes how the residual variance is distributed over temporal frequencies, the PCA eigenvalue spectrum describes how the variance is distributed among orthogonal spatial modes[29]. Therefore, these metrics characterize complementary properties of the corrected data. The leading eigenvalues describe the dominant spatial variance components, whereas the high-index tail represents the progressively weaker spatial variance components. In both, $\Delta\varphi$ and $\Delta\vartheta$, the uncorrected spectrum exhibits larger leading eigenvalues, while the corrected spectra are substantially reduced and largely overlap over the intermediate and high-index ranges. The Zernike- and CNR-corrected spectra show only small differences in the high-index region. Because the datasets were decomposed independently, equal PCA indices do not necessarily correspond to identical spatial eigenvectors. The small crossing between the uncorrected and CNR-corrected curves is therefore interpreted as a redistribution of variance among weak high-order modes, rather than as evidence of greater total noise after CNR correction. Thus, the PCA analysis indicates broadly similar spatial variance distributions with minor differences in the weak modes, whereas the PSD analysis reveals stronger suppression of temporal fluctuations by CNR.

### *4.3 Standard deviation and number of effective independent pixels*

To quantify the effect of the different correction methods on spatial averaging, we analyzed the standard deviation $\sigma(N)$ within the signal region as a function of the number of spatially averaged pixels $N$. For statistically independent noise, spatial averaging reduces the standard deviation according to the expected $1/\sqrt{N}$ scaling. In contrast, spatially correlated noise is less efficiently suppressed by averaging and therefore leads to deviations from this ideal scaling, particularly at larger averaging areas[36]. The corrected and uncorrected datasets were compared with a Poissonian shot-noise reference derived from the experimentally measured photon-count distribution. It is important to distinguish this experimental Poisson reference from the simulated shot noise limit (Fig. 2). While Figure 2b establishes the fundamental multiplexing-dependent sensitivity limit under idealized conditions, the reference used here is calculated directly from the photon-count statistics of the experimental dataset, and therefore accounts for the actual detected intensity of the measurement (see Supplementary Information for further details). As shown in Fig. 4a,b, CNR and CNR+BG provided the largest reduction in the measured standard deviation and followed the shot-noise reference more closely than the Zernike correction over the investigated range of the averaging areas. The small crossings between the corrected curves and the shot-noise reference for $\Delta\varphi$ should not be interpreted as exceeding the shot-noise limit, but rather as statistical deviations within the uncertainty of the reference estimate. For the $M = 24{,}000$ independent realizations used to generate the shot-noise reference, the statistical standard error of the estimated standard deviation was approximately $SE(\sigma)/\sigma \approx 1/\sqrt{2(M-1)} \approx 0.46\%$.

As the measurements are intended for spatially resolved transient imaging, the noise reduction obtained by spatial averaging must be considered together with the associated loss of spatial information. In our analysis, the $N$ pixels are averaged over a square region centered on the field of view, so that the linear dimension of the averaging region increases as $\sqrt{N}$. For the 25 × 25 µm² field of view sampled by 1021 × 1021 pixels, averaging $N = 10^3$ pixels correspond to an approximately 32 × 32-pixel region, or 0.78 × 0.78 µm² in the sample plane. Thus, spatial averaging improves sensitivity by reducing random fluctuations, but at the cost of spatial resolution because each reported value represents an average over a finite area. The purpose of this analysis is therefore to quantify the sensitivity–spatial-resolution trade-off and to determine how efficiently the available spatial sampling contributes to noise reduction in the presence of spatial correlations.

To quantify the departure from ideal independent-pixel averaging, we introduced the effective number of statistically independent pixels, $N_{eff}$ [37,38], as shown in Fig. 4c,d. Such a $N_{eff}$ provides a measure of how much independent spatial information contributes to the averaged signal as the region of interest (ROI) size increases. For statistically independent pixels, $N_{eff} = N$, corresponds to the ideal $1/\sqrt{N}$ reduction in standard deviation. Values of $N_{eff} < N$ indicate that averaging is less efficient than in the independent pixel case, consistent with the presence of spatially correlated noise. Both CNR-based approaches yielded substantially larger $N_{eff}$ values than the uncorrected and Zernike-corrected data, indicating a more effective suppression of spatial correlations. CNR+BG generally provided the closest approach to the ideal $N_{eff} = N$ behavior over the investigated averaged pixel range. The remaining deviations from the ideal relationship indicate that residual correlations and other nonideal noise contributions persist after the correction.

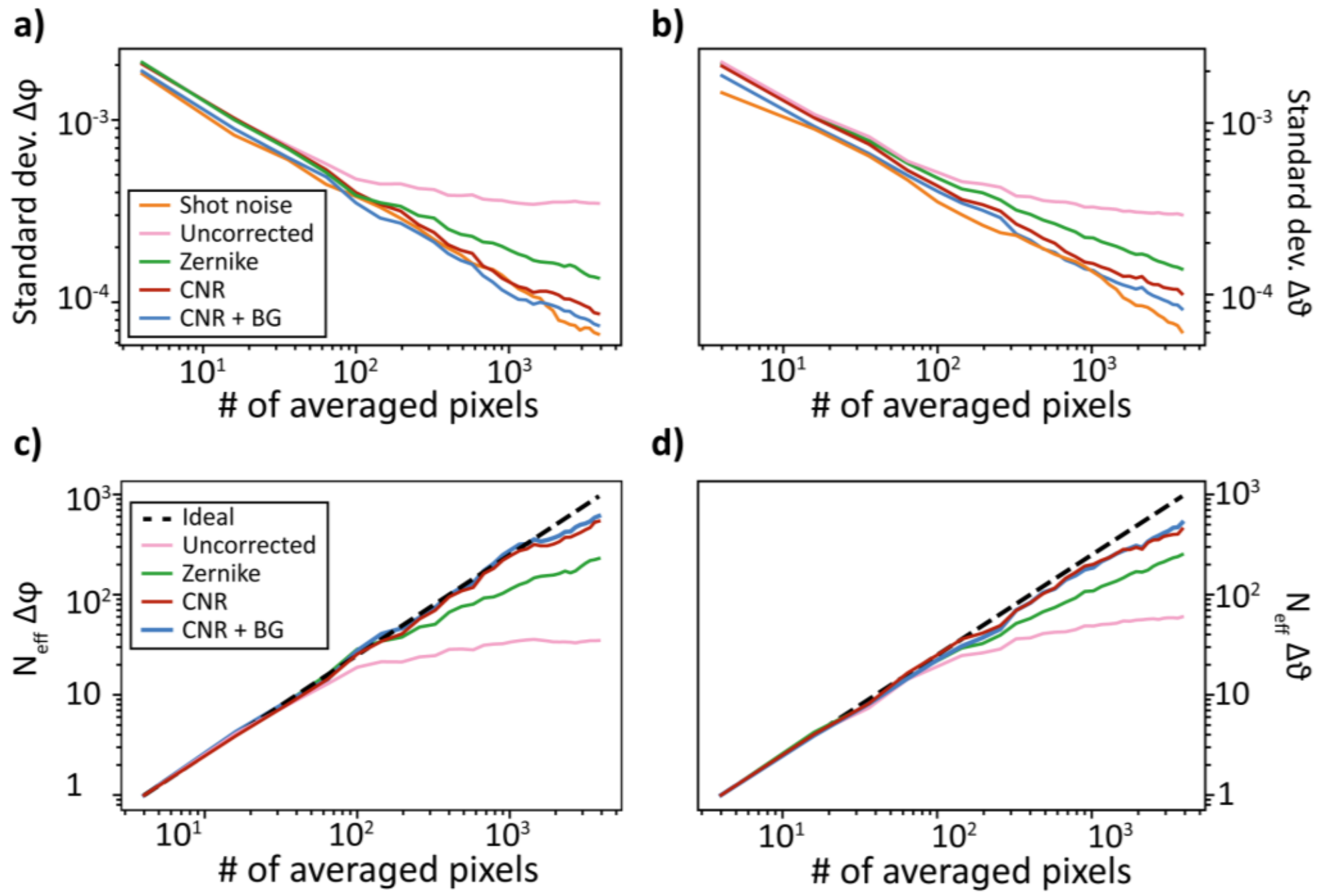

Figure 4: a) and b) Standard deviation as a function of the number of spatially averaged pixels $N$, for $\Delta\varphi$ and $\Delta\vartheta$, respectively. The uncorrected data and the data processed with the different correction methods are compared with a calculated Poissonian shot-noise reference derived from the experimentally measured photon-count distribution. (c,d) Effective number of statistically independent pixels, $N_{eff}$ as a function of the number of spatially averaged pixels $N$ for $\Delta\varphi$ and $\Delta\vartheta$, respectively. The uncorrected data and the corrected datasets are compared with the ideal case of statistically independent pixels $N_{eff} = N$, indicated by the dashed line.

### *4.4 High Sensitivity Transient Holographic Full-Polarization Microscopy*

Finally, the performance of the proposed noise-reduction approach was evaluated on transient measurements acquired in the two-camera configuration shown in Fig. 1a. Two cameras were used, and no optical chopper was employed in the reference arm. As discussed above, this two-camera implementation increases the photon budget of the 1-Ref geometry relative to the multiplexed configurations, with approximately twice the optical intensity incident on each camera compared with the one-camera configuration used for the noise characterization. This results in a different photon and noise regime for the transient measurements.

The pump-probe delay positions were acquired in random order so that signal-free negative-delay measurements were distributed throughout the acquisition, providing representative noise statistics for the self-referenced CNR. The total measurement time was approximately 20 minutes. The pump and probe wavelengths were 490 nm and 540 nm, respectively, with a pump fluence of 13 μJ/cm$^2$. Figure 5 shows the signals $\Delta\varphi$ and $\Delta\vartheta$ measured from a thin methylammonium lead tribromide ($MAPbBr_3$) single crystal. The distinct temporal evolution of these observables reflects their different physical origins: $\Delta\varphi$ is primarily sensitive to the pump-induced refractive-index change associated with the photoexcited charge carrier population, whereas $\Delta\vartheta$ quantifies the transient magneto-optical response induced by a circularly polarized pump (see Section 2.1). Figures 5a and b compare the transient image at three different time steps: before ($t_1$), during ($t_2$), and after the pump arrival ($t_3$), before and after applying the combined correlated noise removal (CNR) and background correction (BG). The corresponding transient signal spatially averaged over the pumped region (16.5 × 16.5 μm$^2$, or 59 × 59 pixels, approximately half the FOV after Fourier filtering), is shown in (c) and (d) as a function of the pump-probe delay for both the corrected and uncorrected data.

Noise reduction was clearly visible in both the spatial images and the averaged transient traces. Applying the combined correction reduced the standard deviation of the pre-time-zero baseline by a factor of 60 in $\Delta\varphi$ and 3 in $\Delta\vartheta$ for the spatially averaged data. Moreover, even when averaging over a 1.4 × 1.4 μm² area (5 × 5 pixels), we observe a reduction in the standard deviation of the pre-time-zero baseline by factors of ≈22 in $\Delta\varphi$ and ≈1.7 in $\Delta\vartheta$, demonstrating the effectiveness of the proposed approach under realistic experimental conditions.

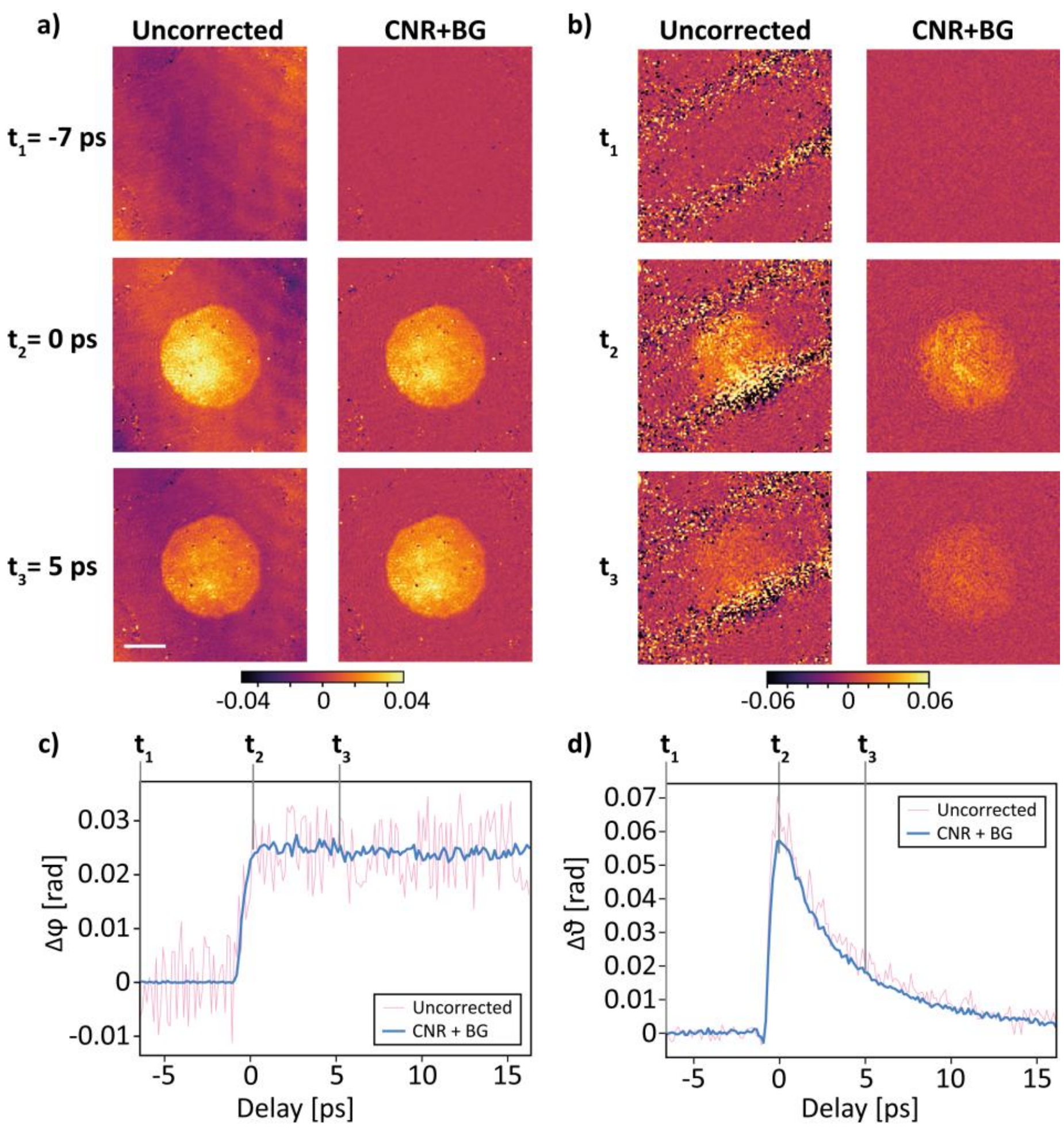

Figure 5: Image sequences at three different time delays before ($t_1$), during ($t_2$), and after the pump arrival ($t_3$), shown without and with noise correction (CNR+BG) for a) $\Delta\varphi$ and b) $\Delta\vartheta$. The scale bar corresponds to 6 μm. Spatially averaged signals are shown in c) and d), demonstrating a noise reduction of the pre-time-zero baseline by factors of 60 for the spatially averaged $\Delta\varphi$ and 3 for $\Delta\vartheta$, respectively. The sample is a $MAPbBr_3$ single crystal excited with a pump fluence of 13 μJ/cm$^2$.

## 5. Conclusions

We systematically investigated the noise performance of transient holographic full-polarization microscopy by comparing three holographic geometries, employing one, two, and four reference beams. The different configurations were evaluated using complementary measures of noise performance, including the standard deviation, power spectral density, principal component analysis, and effective number of statistically independent pixels. Together, these analyses provide a comprehensive characterization of the spatial and temporal properties of the noise and the effectiveness of different correction strategies.

Among the investigated geometries, the single-reference configuration exhibited the most favourable shot-noise-limited SNR for an individual reconstructed carrier. At fixed total photon counts, the raw camera shot-noise variance is independent of the number of reference beams; however, multiplexing distributes the available reference photons among multiple carriers, reducing the signal amplitude of each reconstructed carrier while the camera shot-noise variance remains unchanged.

For transient measurements, the two-camera 1-Ref implementation enables the simultaneous detection of both polarization components, with each component reconstructed from a single interference term, rather than being multiplexed with additional carriers. The 1-Ref geometry therefore avoids the sensitivity penalty associated with distributing the available reference photons among multiple carriers and also avoids the intensity losses associated with the 2D diffraction gratings used for multiplexing.

The combination of self-referenced CNR based on singular value decomposition and background subtraction effectively suppressed both spatially correlated noise and residual smooth background variations. The improved performance of CNR+BG relative to CNR alone is consistent with the removal of the dominant, slowly varying background component prior to the CNR. This preprocessing reduces the variance associated with the dominant correlated background, allowing the CNR procedure to more effectively capture and suppress the remaining correlated fluctuations. Therefore, CNR+BG provided the strongest overall noise reduction among the correction strategies investigated, bringing the experimental noise substantially closer to the Poissonian shot-noise reference. For transient signals spatially averaged over an area of $16.5 \times 16.5$ µm² ($1.4 \times 1.4$ µm$^2$), the noise reduction was quantified from the standard deviation of the pre-time-zero baseline of the transient measurements and reached factors of approximately 3 (1.7) to 60 (22), depending on the detected quantity, thereby substantially improving the sensitivity to weak transient signals.

Overall, the presented combination of single-reference holography and self-referenced numerical noise correction provides an effective route toward high-sensitivity wide-field transient full-polarization microscopy. SVD-based correction requires no separate acquisition of reference or noise-only images, making the approach experimentally straightforward and broadly applicable to wide-field transient microscopy techniques beyond holographic imaging. The demonstrated noise suppression and sensitivity enhancement provide a pathway toward investigating emerging material systems with intrinsically weak signals, including low-dimensional and nanoscale materials, organic semiconductor thin films, and transient responses at low excitation densities.

**Funding:** J.A. Gessner, G. Maliyekkal, T. Buckup and F. Deschler thank the Deutsche Fördergemeinschaft for funding through the Sonderforschungsbereich (SFB) 1249. P. Kollenz, G. May and F. Deschler acknowledge funding from the Deutsche Forschungsgemeinschaft (DFG, German Research Foundation) - GRK 2948 – 508754442. This project was funded by the European Research Council (ERC Starting Grant Agreement no. 852084 - TWIST).

**Acknowledgments**: The authors acknowledge Dr. Franco V. A. Camargo, Prof. Giulio Cerullo, Dr. Federico Visentin, and Dr. Martin Hörmann for their fruitful discussions.

**Disclosures:** The authors declare no conflicts of interest.

**AI Disclosure:** The authors used ChatGPT (GPT-4.1 version) solely for grammar refinement and organizational suggestions. This tool was not used to generate scientific claims, references, experimental data, or conclusions.

**Data availability:** The data underlying the results presented in this study are available upon request.

## 6. References

[1] M. Delor, H. L. Weaver, Q. Yu, and N. S. Ginsberg, *Imaging material functionality through three-dimensional nanoscale tracking of energy flow.* Nature Materials 19, 1, 56–62 (2020). DOI=10.1038/s41563-019-0498-x.

[2] Z. Guo, J. S. Manser, Y. Wan, P. V. Kamat, and L. Huang, *Spatial and temporal imaging of long-range charge transport in perovskite thin films by ultrafast microscopy.* Nature Communications 6, 1, 7471 (2015). DOI=10.1038/ncomms8471.

[3] X. Chen, J. Guo, L. Peng, Q. Wang, S. Jiang, and Y. Li, *Charge transport in organic field-effect transistors.* Materials Today Electronics 6, 100077 (2023). DOI=10.1016/j.mtelec.2023.100077.

[4] C. Schnedermann, J. Sung, R. Pandya, S. D. Verma, R. Y. S. Chen, N. Gauriot, H. M. Bretscher, P. Kukura, and A. Rao, *Ultrafast Tracking of Exciton and Charge Carrier Transport in Optoelectronic Materials on the Nanometer Scale.* The Journal of Physical Chemistry Letters 10, 21, 6727–6733 (2019). DOI=10.1021/acs.jpclett.9b02437.

[5] Martin Hörmann, Federico Visentin, Suman Kumar Chakraborty, Biswajeet Nayak, Prasana Kumar Sahoo, Giulio Cerullo, and Franco V. A. Camargo, *Self-referencing for quasi shot-noise-limited widefield transient microscopy.* Opt. Express 32, 12, 21230–21242 (2024). DOI=10.1364/OE.525581.

[6] R. Wu, Y. Zhang, M. Shahjahan, and E. Harel, *Self-referencing ultrafast wide-field pump-probe microscopy.* Nature Communications 16, 1, 9686 (2025). DOI=10.1038/s41467-025-64680-9.

[7] N. Gross, C. T. Kuhs, B. Ostovar, W.-Y. Chiang, K. S. Wilson, T. S. Volek, Z. M. Faitz, C. C. Carlin, J. A. Dionne, M. T. Zanni, M. Gruebele, S. T. Roberts, S. Link, and C. F. Landes, *Progress and Prospects in Optical Ultrafast Microscopy in the Visible Spectral Region: Transient Absorption and Two-Dimensional Microscopy.* The Journal of Physical Chemistry C 127, 30, 14557–14586 (2023). DOI=10.1021/acs.jpcc.3c02091.

[8] Y. Zhang, B. Shen, T. Wu, J. Zhao, J. C. Jing, P. Wang, K. Sasaki-Capela, W. G. Dunphy, D. Garrett, K. Maslov, W. Wang, and L. V. Wang, *Ultrafast and*

*hypersensitive phase imaging of propagating internodal current flows in myelinated axons and electromagnetic pulses in dielectrics.* Nature Communications 13, 1, 5247 (2022). DOI=10.1038/s41467-022-33002-8.

[9] R. L. McClain and J. C. Wright, *Description of the Role of Shot Noise in Spectroscopic Absorption and Emission Measurements with Photodiode and Photomultiplier Tube Detectors: Information for an Instrumental Analysis Course.* Journal of Chemical Education 91, 9, 1455–1457 (2014). DOI=10.1021/ed4005112.

[10] Hobbs, P. (1991, *Shot noise limited optical measurements at baseband with noisy lasers* 1376. SPIE 1376.

[11] F. Soldevila, P. Clemente, E. Tajahuerce, N. Uribe-Patarroyo, P. Andrés, and J. Lancis, *Computational imaging with a balanced detector.* Scientific Reports 6, 1, 29181 (2016). DOI=10.1038/srep29181.

[12] Nicolas Verrier and Michael Atlan, *Off-axis digital hologram reconstruction: some practical considerations.* Appl. Opt. 50, 34, H136-H146 (2011). DOI=10.1364/AO.50.00H136.

[13] M. Kumar, L. Pensia, and R. Kumar, *Single-shot off-axis digital holographic system with extended field-of-view by using multiplexing method.* Scientific Reports 12, 1, 16462 (2022). DOI=10.1038/s41598-022-20458-3.

[14] G. Coppola and M. A. Ferrara, *Polarization-Sensitive Digital Holographic Imaging for Characterization of Microscopic Samples: Recent Advances and Perspectives.* Applied Sciences 10, 13 (2020). DOI=10.3390/app10134520.

[15] M. Liebel, F. V. A. Camargo, G. Cerullo, and N. F. van Hulst, *Ultrafast Transient Holographic Microscopy.* Nano Letters 21, 4, 1666–1671 (2021). DOI=10.1021/acs.nanolett.0c04416.

[16] M. Hörmann, F. Visentin, A. Zanetta, J. Osmond, G. Grancini, N. F. Hulst, M. Liebel, G. Cerullo, and F. V. A. Camargo, *High-Sensitivity Visualization of Ultrafast Carrier Diffusion by Wide-Field Holographic Microscopy.* Ultrafast Science 3, 32 (2023). DOI=10.34133/ultrafastscience.0032.

[17] M. Liebel, F. V. A. Camargo, G. Cerullo, and N. F. van Hulst, *Widefield phototransient imaging for visualizing 3D motion of resonant particles in scattering environments.* Nanoscale 14, 8, 3062–3068 (2022). DOI=10.1039/d1nr06837g.

[18] M. Hörmann, F. Visentin, J. A. Gessner, P. Kollenz, S. Liu, M. Heindl, F. Deschler, G. Cerullo, and F. V. A. Camargo, *Ultrafast holographic chiroptical microscopy.* Nature Photonics, 20, 592–599 (2026). DOI=10.1038/s41566-025-01824-9.

[19] R. Géneaux, H.-T. Chang, A. M. Schwartzberg, and H. J. B. Marroux, *Source noise suppression in attosecond transient absorption spectroscopy by edge-pixel referencing.* Opt. Express 29, 2, 951–960 (2021). DOI=10.1364/OE.412117.

[20] G. Long, C. Jiang, R. Sabatini, Z. Yang, M. Wei, L. N. Quan, Q. Liang, A. Rasmita, M. Askerka, G. Walters, X. Gong, J. Xing, X. Wen, R. Quintero-Bermudez, H. Yuan, G. Xing, X. R. Wang, D. Song, O. Voznyy, M. Zhang, S. Hoogland, W. Gao, Q. Xiong, and E. H. Sargent, *Spin control in reduced-dimensional chiral perovskites.* Nature Photonics 12, 9, 528–533 (2018). DOI=10.1038/s41566-018-0220-6.

[21] S. A. Bourelle, F. V. A. Camargo, S. Ghosh, T. Neumann, Van de Goor, Tim W. J., R. Shivanna, T. Winkler, G. Cerullo, and F. Deschler, *Optical control of exciton spin*

*dynamics in layered metal halide perovskites via polaronic state formation.* Nature Communications 13, 1, 3320 (2022). DOI=10.1038/s41467-022-30953-w.

[22] J. Zerhoch, S. Bodnar, J. E. Lerpinière, S. Liu, T. Neumann, B. Sergl, M. W. Heindl, A. Shcherbakov, A. Elghandour, R. Klingeler, A. B. Walker, and F. Deschler, *Motional Narrowing Effects in the Excited State Spin Populations of Mn-Doped Hybrid Perovskites.* The Journal of Physical Chemistry Letters 15, 10, 2851–2858 (2024). DOI=10.1021/acs.jpclett.3c03466.

[23] J. Ortega Arroyo and M. Liebel, *The Next Dimension: Digital Holography for 3D Interferometric Scattering.* ACS Nano 19, 44, 38630–38640 (2025). DOI=10.1021/acsnano.5c13224.

[24] A. A. Maznev, T. F. Crimmins, and K. A. Nelson, *How to make femtosecond pulses overlap.* Opt. Lett. 23, 17, 1378–1380 (1998). DOI=10.1364/OL.23.001378.

[25] K. C. Robben and C. M. Cheatum, *Edge-pixel referencing suppresses correlated baseline noise in heterodyned spectroscopies.* The Journal of Chemical Physics 152, 9, 94201 (2020). DOI=10.1063/1.5134987.

[26] Hansol Kim, Youngmo Jeong, Kyoungjae Lee, and Yoonchan Jeong, *Noise robust Zernike phase retrieval via learning based algorithm only with 2-step phase shift measurements.* Opt. Express 31, 19, 30248–30266 (2023). DOI=10.1364/OE.497310.

[27] Zhe Zhang, Donghui Zheng, and Lei Chen, *Analysis and suppression of random noise effect in interferometry on absolute measurement of shift rotation.* Opt. Express 33, 20, 42191–42205 (2025). DOI=10.1364/OE.576700.

[28] B. P. Epps and E. M. Krivitzky, *Singular value decomposition of noisy data: noise filtering.* Experiments in Fluids 60, 8, 126 (2019). DOI=10.1007/s00348-019-2768-4.

[29] I. T. Jolliffe and J. Cadima, *Principal component analysis: a review and recent developments.* Philosophical Transactions of the Royal Society A: Mathematical, Physical and Engineering Sciences 374, 2065, 20150202 (2016). DOI=10.1098/rsta.2015.0202.

[30] Dingnan Deng, Weijuan Qu, Qijian Tang, Wenqi He, and Xiaoli Liu, *Single-shot wavelength-multiplexing for off-axis digital holography with a spectral filter.* Opt. Express 29, 22, 36871–36885 (2021). DOI=10.1364/OE.440980.

[31] J. Á. Picazo-Bueno, S. Ketelhut, J. Schnekenburger, V. Micó, and B. Kemper, *Off-axis digital lensless holographic microscopy based on spatially multiplexed interferometry.* Journal of Biomedical Optics 29, S2, S22715 (2024). DOI=10.1117/1.JBO.29.S2.S22715.

[32] Douglas E. Thornton, Cameron J. Radosevich, Samuel Horst, and Mark F. Spencer, *Achieving the shot-noise limit using experimental multi-shot digital holography data.* Opt. Express 29, 6, 9599–9617 (2021). DOI=10.1364/OE.418948.

[33] F. Crisafi, V. Kumar, T. Scopigno, M. Marangoni, G. Cerullo, and D. Polli, *In-line balanced detection stimulated Raman scattering microscopy.* Scientific Reports 7, 1, 10745 (2017). DOI=10.1038/s41598-017-09839-1.

[34] P. Welch, *The use of fast Fourier transform for the estimation of power spectra: A method based on time averaging over short, modified periodograms.* IEEE Transactions on Audio and Electroacoustics 15, 2, 70–73 (1967). DOI=10.1109/TAU.1967.1161901.

[35] Oppenheim, A. V., Schafer, R. W., and Buck, J. R. (1999, *Discrete-time Signal Processing*. Prentice Hall International Editions Series. Prentice Hall.

[36] T. Tsukui, S. Iguchi, I. Mitsuhashi, and K. Tadaki, *Estimating the statistical uncertainty due to spatially correlated noise in interferometric images.* Journal of Astronomical Telescopes, Instruments, and Systems 9, 1, 18001 (2023). DOI=10.1117/1.JATIS.9.1.018001.

[37] Ronny Vallejos and Felipe Osorio, *Effective sample size of spatial process models.* Spatial Statistics 9, 66–92 (2014). DOI=10.1016/j.spasta.2014.03.003.

[38] Christel Faes, Geert Molenberghs, Marc Aerts, Geert Verbeke, and Michael G. Kenward, *The Effective Sample Size and an Alternative Small-Sample Degrees-of-Freedom Method.* The American Statistician 63, 4, 389–399 (2009). DOI=10.1198/tast.2009.08196.